\documentclass[12pt]{article}
\usepackage{epsfig}
\begin{document}

\begin {center}
{\bf The $D\pi$ S-wave}

\vskip 4mm {D.\ V.\ Bugg\footnote{email:
david.bugg@stfc.ac.uk} \\[2mm]
{\normalsize\it Queen Mary, University of
London, London E1\,4NS, UK} \\[3mm]}
\end {center}
\date{\today}

\begin{abstract}
\noindent
A pole in the $D\pi$ S-wave analogous to the $\sigma$ and $\kappa$ 
is predicted at
${\rm {M}} - i\Gamma /2 = 2098 \pm 40 - i(260 \pm 40)$ MeV.
The main objective of this paper is to provide formulae for fitting it 
to data.

\vskip  2mm


\end{abstract}

\vskip  2mm

There are well-known $\sigma$ and $\kappa$ poles in $\pi \pi$ and $K\pi$.
The objective of this paper is to predict analogous poles in $D\pi$ and $B\pi$
isospin $1/2$ S-waves and provide appropriate parametrisations of them.
It has not been observed experimentally yet.
There are earlier predictions of the poles in $D\pi$ and $B\pi$ S-waves by
van Beveren and Rupp \cite {Beveren}.
Their result for $D\pi$ agrees fairly closely with that found here.
For $B\pi$ they predict the $^3P_0$ $b\bar n$ state close to the $B\pi$
threshold. 
Guo et al. give similar predictions for poles in $D\pi$ and $B\pi$ 
but with widths roughly a factor 2 smaller than those found here \cite {Zou}.
Another recent paper by Guo with other authors discusses the
$D\pi$ and $B\pi$ scattering lengths and related pole positions \cite {Guo}.

The existence of these poles and their parameters are of interest to understanding
Chiral Symmetry Breaking and its possible relation to Confinement.
These could be separate phase transitions, i.e. condensates.
On the other hand it is possible that they are two facets of a single phase
transition.
What will emerge from the discussion here is that the predicted pole near the
$D\pi$ threshold overlaps significantly with the lowest $^3P_0$ $c\bar n$ state.
How or whether they mix and possibly coalesce is of fundamental interest to
understanding how confinement works.

Let us begin by reviewing briefly the facts about $\sigma$ and $\kappa$.
Caprini et al. predict the $\sigma$ pole at
$441 ^{+16}_{-8} - i272 ^{+8}_{-12.5}$ MeV \cite {Caprini}.
It appears as a peak in E791 data for $D \to 3\pi$
\cite {Aitala} at $\sim 478$ MeV and, more prominently, in BES2 data on
$J/\Psi \to \omega \pi ^+\pi ^-$ at $\sim 500$ MeV \cite {Ablikim}.
Pole positions from these and other data depend somewhat on formulae used to 
fit them.

Descotes-Genon et al. predict a similar $\kappa$ pole in $K\pi$
elastic scattering at $658 \pm 13 - i(279 \pm 12)$ MeV
\cite {Descotes}
The corresponding peak is clearly visible in $D\to K\pi \pi$ data of E791
\cite {Aitala2} \cite {E791} and less clearly in BES2 data for $J/\Psi \to
KK\pi \pi$ \cite {Abl2} \cite {DVB}.
Again experimental values are somewhat higher in mass, but
strongly dependent on fitting formulae.
Most analyses of data do not include the $s$-dependence
required by Chiral Symmetry Breaking, hence confusing the
experimental situation.
This needs to be remedied and a prime objective of the present paper is to provide
simple formulae and discuss their foundations.
Algebra will be presented for the $D\pi$ S-wave and the
resulting pole will be called $\partial$ (dabba) as a
shorthand.
Formulae carry over trivially to $\pi \pi$, $K\pi$ and $B\pi$.

For a resonance, the elastic scattering amplitude is
\begin {eqnarray}
T(s) &=& \frac {M\Gamma (s)}{M^2 - s - iM\Gamma (s)} \\
T^{-1}(s) &=& \frac {M^2 - s}{M\Gamma (s)} - i
                = \cot \delta  - i,
\end {eqnarray}
where $\delta$ is the elastic phase shift.
There are four key points.
\begin {itemize}
\item {There is an Adler zero at $s_A = m^2_D - 0.5m^2_\pi $,
   where $m_D$ and $m_\pi$ are masses of D and $\pi$; it originates from
   Chiral Symmetry Breaking.}
\item {Weinberg \cite {Weinberg} predicts the $I=1/2$ scattering
   length $a = (1/k)\tan \delta $ at threshold:
\begin {equation}
a = \frac {0.22/m_\pi}{1 + m_\pi /m_D},
\end {equation}
where $k$ is centre of mass momentum.
}
\item {This scattering length is small, so pions are
`soft', i.e. interact weakly.
Hence the numerator of $T(s)/\rho(s)$ is not constant as in most resonances, 
but approximately linear with $s$ over the range relevant for the pole.
Here $\rho(s)$ is Lorentz invariant $D\pi$ phase space. }
\item {
For the $\sigma$, $M \simeq 0.9 - 1.0$ GeV; for the $\kappa$,
$M \simeq 3.3$ GeV, well above $K_0(1430)$.
This progression is likely to increase with mass of the
spectator $D$, with the result that $M$ for the $\partial$ in Eq. (1) is likely 
to be far above that of the $^3P_0$ $c\bar n$ state.
The $K\pi$ phase shift reaches only $\sim 70^\circ$ at 1.4 GeV.
Non-linearity in $\Gamma (s)$ might prevent
the elastic $D\pi$ phase shift ever reaching $90^\circ$.
However, there is still a pole near threshold associated with Chiral Symmetry
Breaking.
This confusing point will be discussed below.}
\end {itemize}

Steps in the algebra are to write
\begin {equation}
\Gamma  = B(s)(s - s_A) \rho (s) ,
\end {equation}
where $\rho$ is $D\pi$ phase space; $B(s)$ is
constant to first approximation.
Next, Eq. (2) may be reparametrised as
\begin {equation}
\rho T^{-1} = \frac
{1 - \beta (s - s_{thr})}
{B'(s)(s - s_A)} -i\rho = \rho \cot \delta - i\rho ,
\end {equation}
where $\beta \propto 1/M^2$ is small for $D\pi$; $ s_{thr} = (m_D + m_\pi)^2$
is the value of $s$ at threshold.
This equation gives the relation to Weinberg's prediction
for the scattering length.
Since  $\rho = 2k/\sqrt {s}$,
\begin {equation}
a = 2B'(s_{thr})(s_{thr} - s_A)/\sqrt {s_{thr}}.
\end {equation}
This expression determines $T$ except for
the small parameter $\beta$ and possible
deviations of $B'(s)$ from a constant.
To take account of this non-linearity, one can write
\begin {equation}
B'(s) = b \exp [-\alpha (s - s_{thr})],
\end {equation}
where $b$ is a constant and the exponential prevents $B'(s)$
from  diverging at large $s$.
Essentially, the exponential provides a
convergent power series for $B'(s)$ near threshold.
Inverting (5),
\begin {equation}
T =\frac {B'(s)(s - s_A)\rho}
{1 - \beta (s - s_{thr}) - iB'(s)(s - s_A)\rho}.
\end {equation}

For $D_0(2400)$, Eqs. (1) and (4) can be used with
\begin {eqnarray}
B(s) &=& b'\exp (-2\alpha 'k^2) \\
\alpha ' &=& \frac {1}{6}\left( \frac {R(fm)}{\hbar c} \right),
\end {eqnarray}
$\hbar c = 0.19732$ (GeV/c)$^{-2}$,
corresponding to a Gaussian source.
With $R$ in the range 0.5--0.7 fm for the overlap of the $D_0$ with its
decay products $D$ and $\pi$, $\alpha ' $ is expected in the range
1--2 (GeV/c)$^{-2}$.

Some approximations in the algebra require brief discussion.
Firstly, inelasticity is neglected.
The first obvious inelastic threshold is $D\eta$ at 2416 MeV,
but is predicted to couple weakly to  $J^P = 0^+$ \cite {Tornqvist},
as is confirmed experimentally for $\kappa \to K\eta$ and $\sigma \to \eta \eta$.
The $D\eta '$ threshold is far above the relevant mass range.
The $D\pi \sigma$ threshold is  high in mass, but the $D_sK$ channel could
contribute from 2460 MeV.
Up to the mass of $D_0(2400)$, neglect of inelasticity should be reliable.

A more significant point is that there is a weak cusp in
$B(s)$ at the $D\pi$ threshold, i.e. a discontinuity in the slope of 
$B'$ at threshold.
For the $\sigma$, this affects the scattering length by
$\sim 30\%$, but only by $10\%$ for $K\pi$ and the effect
is likely to be still smaller for $D\pi$.
The cusp can be taken into account exactly using a dispersion integral 
\cite {sigpole}, but this is somewhat laborious;
it is also unnecessary in view of larger uncertainties to be discussed shortly.

With these approximations, column 4 of Table 1 shows an initial set of predictions 
for the $\partial$ pole (and for $B\pi$)
for some values of $\beta$ (i.e. the term
$\beta (s - s_{thr})$ in the numerator of $T$) and
$\alpha$ (the coefficient in the exponential).
The value of $B'$ at threshold, hence the value of $b$ in  Eq. (7), is taken
from Weinberg's prediction of the scattering length, Eq. (6). 
For the $\kappa$, $\beta = 0.25$ \cite {kappa}.
The parameter $M$ of Eq. (1) (not to be confused with the
pole mass) increases strongly from $\sigma$ to $\kappa$ and
is likely to be larger for $\partial$, so $\beta $ is likely
to be $<0.25$.
The parameter $\alpha \sim 0.4$ GeV$^{-2}$ for $\sigma$ \cite {sigpole}
and $0.25$ for $\kappa$ \cite {kappa} and is likely to be smaller for
$\partial$.
The conclusion from Table 1 is that the $\partial$ has a
pole mass $\sim 60$ MeV above the $D\pi$ threshold (2009 MeV)

\begin {table} [htb]
\begin {center}
\begin {tabular}{ccccc}
System & $\beta$ (GeV$^{-2}$) & $\alpha$ (GeV$^{-2})$& Pole (MeV) & Pole ' (MeV)
 \\
\hline
$D^+\pi ^-$ & 0.2 & 0.0 & 2113 -i202 & 2136 - i221\\
            &     & 0.1 & 2108 -i216 & 2131 - i240\\
            &     & 0.2 & 2099 -i231 & 2121 - i259\\
            &     & 0.3 & 2087 -i246 & 2106 - i279\\
            & 0.1 & 0.0 & 2105 -i216 & 2128 - i239 \\
            &     & 0.3 & 2069 -i256 & 2082 - i291 \\
            & 0.01& 0.0 & 2097 -i228 & 2117 - i256\\
            &     & 0.3 & 2052 -i263 & 2060 - i299\\\hline
$B^+\pi ^-$ & 0.2 & 0.0 & 5525 -i148 \\
            &     & 0.1 & 5529 -i176 \\
            &     & 0.2 & 5526 -i218 \\
            &     & 0.3 & 5487 -i273 \\
            & 0.1 & 0.0 & 5517 -i177 \\
            &     & 0.2 & 5477 -i247 \\
            & 0.01& 0.0 & 5504 -i206 \\
            &     & 0.2 & 5433 -i256 \\\hline
\end{tabular}
\caption {Predicted pole positions for $D^-\pi ^+$ and
$B^- \pi ^+$; column 4 ignores the estimated contribution to the
scattering length from $D_0(2400)$ and column 5 includes it}
\end{center}
\end {table}

However, there is a further point. 
Weinberg's prediction for the scattering length applies to the full
$D\pi$ S-wave, i.e. to the coherent sum of $\partial$, $D_0(2400)$ and small
contributions from even higher resonances (which are likely to be inelastic).
In $\pi \pi$ elastic scattering, Cern-Munich data \cite {hyams}
show clearly that phases of $\sigma$ and $f_0(980)$ add near 1 GeV.
The standard prescription is to multiply $S$-matrices $S = \exp (2i \delta)$ so as
to satisfy unitarity; phases then add.
This will be assumed for the overlap of $\partial$ and $D_0(2400)$.

There are two published sets of data on $D_0(2400)$ from Belle \cite {Abe} and 
Focus \cite {Focus}. 
Belle data show a clear peak above modest (50\%) experimental background.
Belle fit the data with a resonance with width proportional to $k$ (centre of mass
momentum) and find a mass of 2308 MeV and $\Gamma = 276 $ MeV on resonance.
Focus data contain a strong combinatoric background at low mass and find
$M = 2407$ MeV, $\Gamma = 240$ MeV, using a constant width.
A recent preprint from Babar, almost simultaneous with the present article,
provides data with smaller experimental background and a resonance mass of 
$2297 \pm 8 \pm 5 \pm 19$ MeV  and a width of $273 \pm 12 \pm 17 \pm 45$ MeV
\cite {Babar_new}.
These data do exhibit a shoulder at $\sim 2100$ MeV, very similar to that
predicted here.
It has been fitted as a coherent `background' in the $0^+$ amplitude
and an incoherent P-wave background. 

Belle's parameters have been adopted here using  central values of $M$ and 
$\Gamma$  but using Eqs. (4) and (6).
Parameters of Babar will give results in agreement well within the errors.
The contribution of $D_0(2400)$ to the scattering
length is 17.2\% of Weinberg's prediction if no form factor is included 
(i.e. $\alpha = 0$).
It rises to 20.9\% using $\alpha = 0.15$ GeV$^{-2}$ in the form factor.
Some form factor is needed to account for the size of $D_0(2400)$, but on the
other hand Focus parameters predict a smaller contribution to the scattering
length. 
As a compromise, a contribution of 17.2\% will be used for $D_0(2400)$.
The last column of Table 1 shows the quite small effect on parameters of the
$\partial$. 
Its width increases because of the reduced scattering length.

There remains the important question of how the $\partial$ and $D_0(2400)$ 
should be fitted to production data.
The denominator of both resonances should be the same as in elastic scattering
\cite {Watson}.
However, numerators can be quite different to elastic scattering.
Data for $\sigma$ and $f_0(980)$ rule out the hypothesis of adding their
phases as in elastic scattering \cite {Extended}.
The procedure which has worked successfully for $\sigma$ and $\kappa$ is the
standard isobar model, where each resonance is fitted with a complex coupling
constant $\Lambda$ replacing the numerator of Eq. (8).
This reproduces the low mass peaks observed in BES2 and E791 data.
However, a cautionary comment is that for $\bar pp \to 3\pi ^0$ data, 
although there is a low mass $\sigma$ peak, an accurate fit requires 
using a numerator $\Lambda _1 + \Lambda _2(s - s_{thr})$, where $\Lambda$ are
fitted complex complex constants  \cite {f01370}.
This could happen also for $D\pi$ data, since it is presently
unclear why and how the Adler zero disappears from the
numerator between elastic scattering and production reactions.

\begin{figure}[htb]
\begin{center}
\vskip -12mm
\epsfig{file=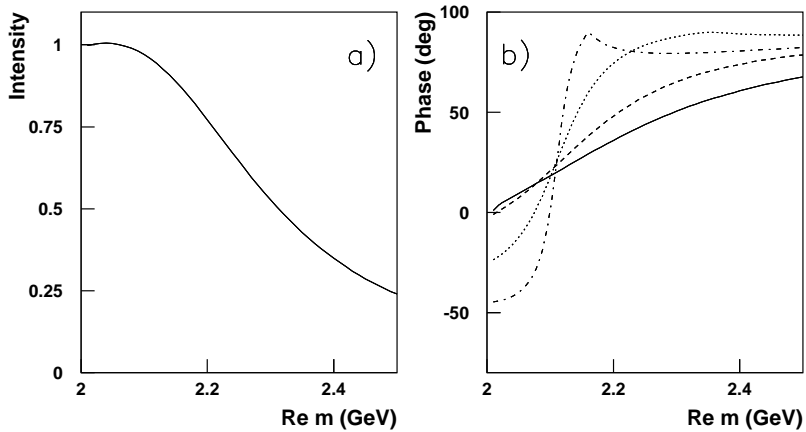,width=12cm}
\vskip -6mm
\caption
{(a) the intensity of the pole on the real $s$-axis (full curve) for
$\beta = 0.1$, $\alpha = 0.15$ GeV$^{-2}$;
it is normalised to 1 at its peak;
(b) the phase shift for $Im~m = 0$ (full curve), $-0.1$ GeV (dashed),
$-0.2$ GeV (dotted) and $-0.245$ GeV (chain curve)}
\end{center}
\end{figure}

Fig. 1(a) shows the line-shape predicted for the $\partial$ if the constant
numerator is used.
It peaks close to threshold and falls to half-height just above 2.3 GeV.
This signal would be obscured in Focus data by their combinatorial background.
Belle data could accomodate a modest contribution from the $\partial$,
but it is not possible to refit their data without knowing their acceptance as 
a function of mass.

A point causing confusion is the possibility of a pole close to threshold despite
the fact that the elastic phase shift (on the real $s$-axis) does not reach
$90^\circ$.
On the real $s$-axis, the phase shift is constrained by
unitarity to start from zero at threshold.
However, off the real $s$-axis the zero of phase can be
displaced to lower values of $s$.
The amplitude describing the resonance is an analytic
function of $s$ and satisfies the Cauchy-Riemann relations:
$d(Re~T)/d(Re~s) = d(Im~T)/d(Im~s)$ and
$d(Im~T)/d(Re~s) = -d(Re~T)/d(Im~s)$.
Because the width is strongly $s$-dependent on the
real $s$-axis, there is a rapid variation of the phase
of the amplitude as one goes off the real $s$-axis into
the lower part of the complex plane.
A valuable discussion of this point is to be found in a paper of 0ller \cite
{Oller}.
 
Fig. 1(b) illustrates what happens. 
It shows the phase shift of the $\partial$
for values of $Im~m = -0.1$, $-0.15$, $-0.2$ and $-0.245$ GeV as full,
dashed, dotted and chain curves.
Phases near $Re~m = 2.5$ GeV change by small amounts as one moves into the 
complex plane.
However, for $Re~m$ near threshold, the phase moves strongly negative,
i.e. there is a phase rotation of the amplitude. 
Near the pole (chain curve),
the phase variation approaches $180^\circ$ as one passes the pole.

For the record, values of the corresponding pole for $D-K$ and $B-K$ are
recorded in Table 2. 
For $D-K$ the pole is very wide, of order 1 GeV. 
With this width, it will be very hard to observe.
The large width arises from the fact that the Adler zero at $m^2_D - 0.5m^2_K$
is much further from the physical region than for $D_\pi$, resulting in a 
smaller gradient of the amplitude as a function of $s$.
The fact that $f_K = 1.193 f_\pi$ also makes the scattering length smaller
and the pole wider.
The mass of the pole varies rather strongly with $\alpha$ and $\beta$.
For $B-K$, the pole position is even less stable in mass.
Its width is somewhat smaller, but again sensitive to $\alpha$ and $\beta$.
\begin {table} [htb]
\begin {center}
\begin {tabular}{cccc}
$\beta$ (GeV$^{-2}$) & $\alpha$ (GeV$^{-2})$& $D-K$ Pole (MeV) & $B-K$ pole (MeV) \\
\hline
 0.2 & 0.0 & 2597 -i421 & 5958 - i226\\
     & 0.1 & 2591 -i547 & 6048 - i293\\
     & 0.2 & 2428 -i710 & 6202 - i205\\
     & 0.3 & 2225 -i671 & 6048 - i293\\
 0.1 & 0.0 & 2538 -i515 & 5932 - i335 \\
     & 0.1 & 2439 -i621 & 5850 - i541 \\
     & 0.2 & 2278 -i641 & 5544 - i496\\
     & 0.3 & 2158 -i583 & 5450 - i340\\
0.01 & 0.0 & 2445 -i592 & 5839 - I454\\
     & 0.1 & 2304 -i626 & 5620 - I493\\
     & 0.2 & 2181 -i585 & 5482 - I369 \\
     & 0.3 & 2103 -i518 & 5439 - I248 \\\hline
\end{tabular}
\caption {Predicted pole positions for $D^-K^+$ and
$B^-  K^+$.}
\end{center}
\end {table}

It is rather important to make the attempt to locate the
$\partial$.
Chiral Symmetry Breaking is a feature of the pion and should
be unaffected by the particle with which is it is produced:
$\pi$, $K$, $D$ or $B$.
The origin of Chiral Symmetry Breaking has been presented
in many places, of which examples are the work of
Bicudo and Ribiero \cite {Bicudo}, Bicudo et al. \cite
{Bicudo2} and Roberts et al. \cite {Roberts}.
Van Beveren and Rupp suggest that major mixing arises
between the $\partial$ and $D_0(2400)$ \cite {Rupp}.
This could increase the width of the $\partial$ substantially, rendering
it invisible. 
This would indicate a merging of Chiral Symmetry Breaking with Confinement.
This is a fundamentally important issue.
Experimentally, possible mixing between $\partial$ and $D_0(2400)$ may be treated
using the formalism of Anisovich et al. \cite {Andrei}.

\end{document}